\documentclass[doublecol]{epl2} 
\usepackage{appendix}
\usepackage{amsmath}

\title{Quantifying long-range correlations in complex networks\\ 
beyond nearest neighbors}

\shorttitle{Quantifying long-range correlations in complex networks}

\author{Diego Rybski$^{1}$, Hern\'an D. Rozenfeld$^{2,3}$, 
	and J\"urgen P. Kropp$^{1}$}

\shortauthor{D. Rybski \etal}

\institute{                    
 \inst{1} Potsdam Institute for Climate Impact Research - 
 	14412 Potsdam, Germany, EU\\
 \inst{2} Levich Institute, City College of New York - 
 	New York, NY 10031, USA\\
 \inst{3} Division of Natural Sciences, College of Mount Saint Vincent - 
 	Riverdale, NY 10471, USA	
}

\date{\today}

\abstract{
We propose a fluctuation analysis to quantify 
spatial correlations in complex networks. 
The approach considers the sequences of degrees along shortest paths 
in the networks and quantifies the fluctuations in analogy to time series. 
In this work, the Barabasi-Albert (BA) model, 
the Cayley tree at the percolation transition,
a fractal network model, 
and examples of real-world networks are studied. 
While the fluctuation functions for the BA model show exponential decay, 
in the case of the Cayley tree and 
the fractal network model the fluctuation functions display a 
power-law behavior. 
The fractal network model comprises long-range anti-correlations. 
The results suggest that the fluctuation exponent provides complementary 
information to the fractal dimension.
}

\begin{document}

\maketitle

\section{Introduction}

Networks, consisting of simple elements, its nodes and links, 
can display complex properties, such as a broad degree distribution, 
clustering, modularity, and many others.
This work focuses on the correlation between node degrees -- the number of 
links attached to a node. Degree correlations, measuring the likelihood
that nodes of a given degree are connected, help to explain important 
features of complex networks that are beyond the degree distribution or 
clustering. For instance, it has been found that many networks, 
such as the coauthorship, film actor (IMDb -- Internet Movie Database), 
and company directors networks display assortative mixing between degrees, 
indicating that nodes of like degree tend to be connected. 
On the other hand, the Internet (autonomous system), 
the WWW, and some biological networks, 
exhibit disassortative mixing, where there is a high tendency for 
high degree nodes (hubs) to be connected to low degree nodes 
\cite{NewmanM2002,NewmanMEJ2003}.

Different ways to quantify degree correlations are defined and used in the 
literature. 
One measure is the Pearson correlation coefficient of all linked 
pairs of nodes \cite{NewmanM2002} which derives from the probability 
distribution $p(k_1,k_2)$ that two nodes of degree $k_1$ and $k_2$ are connected 
through a link~\cite{PastorSatorrasVV2001,Maslov2002}. 
Another measure of 
degree correlations is the average degree of neighbors of a degree-$k$ 
node~\cite{PastorSatorrasVV2001}. This measure may also be obtained as a 
particular case of the matrix $p(k_1,k_2)$, and classifies networks into 
assortative (disassortative) when this quantity increases (decreases) with $k$. 
Furthermore, it has been shown that disassortativity reflected in $p(k_1,k_2)$ 
is a tightly related property to fractality in complex networks 
\cite{SongHM2006,GallosSM2008}.

The different measures of degree correlations consider only correlations
between nearest neighbor nodes, i.e. only correlations between the 
degrees of nodes at distance~$1$, but not further. 
Because of this limitation, these measures fail to capture much of the rich 
topological information of the network.
In this work we introduce an approach that extends the idea of 
degree correlations to larger distances. 
An immediate extension of previous methods could consist on simply considering 
the correlation coefficient from nearest neighbors to the second, the third, 
\ldots $d$-th neighbors.
However, in complex networks the direct calculation of such a 
function is not feasible. 
Here we introduce a \emph{fluctuation analysis} 
that overcomes this issue. 
We study the fluctuations of the degree along shortest paths 
between two nodes and consider the distance between nodes analogous to 
time in time-series analysis.
Our approach can be adapted to study many topological and dynamical 
correlation properties in networks, such as node activity, node weights, 
time of node addition, etc. 
In this work we introduce the formalism and focus on degree correlations, 
leaving other properties for further investigation.

\section{Fluctuation analysis}

\begin{figure}
\onefigure[width=1.0\columnwidth]{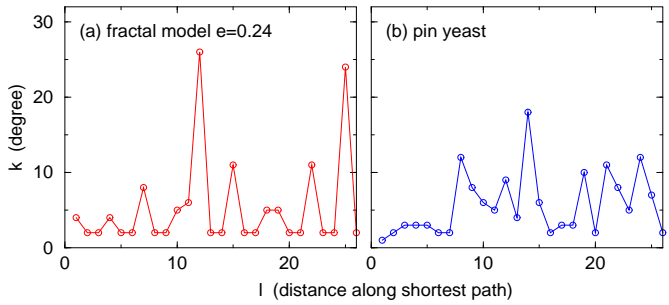}
\caption{
Examples of degree sequences along shortest paths. 
For arbitrary shortest paths we plot the degree of the nodes versus the 
distance for 
(a) the fractal network model with $e=0.24$ ($\alpha_k\approx -0.65$) and 
(b) the pin yeast network ($\alpha_k\simeq -0.53$). 
}
\label{fig:kspex}
\end{figure}

In order to quantify long-range degree (anti-) correlations at distances 
larger than nearest neighbors, we propose a fluctuation analysis. 
The method consists of the following steps:
\begin{enumerate}
\item 
Find the shortest path between all pairs of nodes in the network 
(if the shortest path between a pair of nodes is not unique, 
then we consider an arbitrary one), 
see Fig.~\ref{fig:kspex}.
\item \label{enum:average}
Consider all shortest paths of length $d$ and calculate the average
$K_d = \langle k_l \rangle$ of the sequence ($k_l$) of node degrees 
along each of these paths of length $d$, where $1\le l\le (d+1)$.
\item
Repeat the previous step for all possible values of $d$ in the network.
\item
Calculate the fluctuation function $F(d)=\sigma(K_d|d)$, 
which is the conditional standard deviation of the $K_d$ 
at distance~$d$, i.e. the standard deviation of the 
averages $K_d$ for those paths of length $d$.
\end{enumerate}

The fluctuation function describes the correlations of
node degrees along shortest paths
\cite{PengBGHSSS1992,KoscielnyBundeBHRGS98,KoutsoyiannisD06,KantelhardtJW2010}. 
When an uncorrelated time series is partitioned into segments of size~$s$, 
the standard deviation of the segments averages decays as 
$\sigma\sim s^{-1/2}$.

Because the overall distance between nodes in a network is short 
(compared to the time span of time series), 
we do not partition the degree sequences into segments. 
Instead, we consider all shortest path of any length~$d$. 
Thus, if the covariance, $C(d)$, between the degree of nodes at distance~$d$ 
scales as 
$C(d)\sim\langle (k_i-\langle k\rangle)(k_j-\langle k\rangle)|d\rangle\sim d^{-\gamma}$, 
where $\langle k \rangle$ is the average degree of the network, 
then for the fluctuation function we expect
\begin{equation}
F(d)\sim d^{\alpha_k}
\, ,
\label{eq:Fdsimda}
\end{equation}
where $\alpha_k=-\gamma/2$.
Notice that because the average (calculated in step~\ref{enum:average}) 
involves a division by~$d$, the degree fluctuation exponent~$\alpha_k$ 
differs by 1 from the usual Hurst-like 
exponent~$\alpha$ \cite{KantelhardtKRHB01} of time series analysis: 
$\alpha_k=\alpha-1$.
For asymptotical $\alpha_k=-1/2$ ($\alpha=1/2$) the degrees are 
uncorrelated.

Starting from the average degree along a shortest path, 
$K_d=\frac{1}{d+1}\sum_{j=1}^{d+1}k_j$, 
we can express the variance as follows \cite{KantelhardtKRHB01}:
\begin{equation}
\sigma^2(K_d|d) = \langle (K_d-\langle K_d\rangle)^2\rangle
\, ,
\end{equation}
where $\langle K_d \rangle$ is the average over all values of $K_d$.
Therefore, assuming $\langle K_d\rangle\simeq \langle k\rangle$, we find
\begin{eqnarray}
\sigma^2(K_d|d) &\simeq& \frac{1}{(d+1)^2} 
 \left\langle \sum_{j=1}^{d+1}(k_j-\langle k\rangle)^2\right\rangle
 \nonumber\\
 & & +\frac{1}{(d+1)^2}\nonumber\left\langle 
 \sum_{j\ne l}^{j,l\le d+1}\left[(k_j-\langle k\rangle)(k_l-\langle k\rangle)\right]\right\rangle\nonumber\\
 &=& \frac{1}{(d+1)}\left\langle(k_j-\langle k\rangle)^2\right\rangle\nonumber\\
 & & +\frac{1}{(d+1)^2} \sum_{j\ne l}^{j,l\le d+1} \!\!\! C(|j-l|)\nonumber\\
 &=& \frac{1}{(d+1)}\left\langle(k_j-\langle k\rangle)^2\right\rangle\nonumber\\
 & & +\frac{2}{(d+1)^2} \sum_{j=1}^{d} (d+1-j) C(j) \, .\nonumber
\end{eqnarray}
In the case of (positive) long-range correlations with 
correlation exponent~$\gamma$, 
the second term dominates. 
In the limit of large~$d$ we can integrate the sum, 
leading to the approximation:
\begin{eqnarray}
\sigma^2(K_d|d) &\sim& \frac{2}{(d+1)^2} d^{2-\gamma}\nonumber\\
\sigma(K_d|d) &\sim& d^{-\gamma/2} \, .
\end{eqnarray}
With Eq.~(\ref{eq:Fdsimda}) and $F(d)=\sigma(K_d|d)$ we obtain 
\begin{equation}
\alpha_k=-\gamma/2
\label{eq:kg2}
\, .
\end{equation}

While positive long-range correlations are characterized by 
fluctuation exponents~$-1/2<\alpha_k<0$ \cite{KantelhardtKRHB01}, 
negative long-range correlations (long-range anti-correlations) 
are characterized by fluctuation exponents~$-1<\alpha_k<-1/2$ 
\cite{PengMHHSG1993,BaharKNRRWBM01,MaBBGYI2010}.
For the former, the covariance scales as $C(d)\sim d^{-\gamma}$ and 
for the latter we assume $C(d)\sim -(d^{-\gamma})$.

\section{Results}

\subsection{Barabasi-Albert model}

First we investigate the Barabasi-Albert (BA) model which is based 
on preferential attachment and has been introduced to generate 
scale-free networks \cite{BarabasiA1999} with 
power-law degree distribution~$p(k)\sim k^{-3}$.
The BA model consists of subsequently adding nodes
to the network and linking them to $m$ already existing nodes 
which are randomly chosen with a probability proportional to their degree.
In this model, the parameter $m$ denotes the number of links 
attached to a new node. Therefore, the larger the $m$, the more cycles are 
generated in the network. In addition, when $m$ is large
the overall diameter of the network -- the longest 
shortest path between all pairs of nodes --  decreases 
due to the increased number of paths between any pair of nodes.

\begin{figure}
\onefigure[width=1.0\columnwidth]{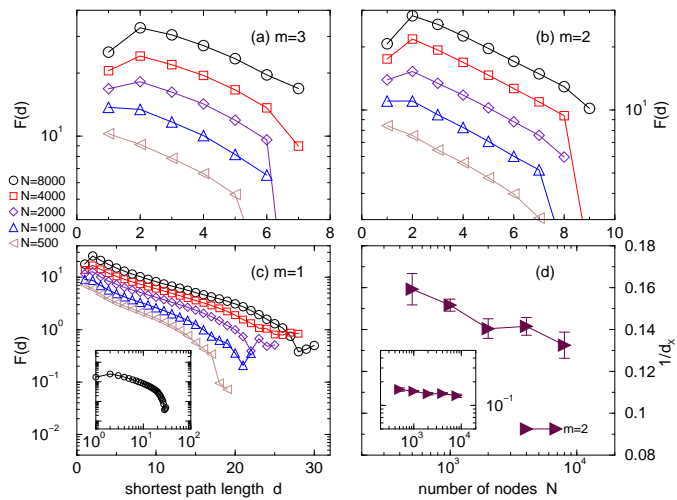}
\caption{
Degree fluctuation functions for the BA model.
The fluctuation functions, $F(d)$, are plotted against the length of the 
shortest paths~$d$, for (a)~$m=3$, (b)~$m=2$, and (c)~$m=1$. 
The panels~(a-c) show $F(d)$ for networks consisting of 
$500$, $1000$, $2000$, $4000$, and $8000$ nodes (from bottom to top). 
Panel~(d) shows for $m=2$ the slopes of exponential fits applied to $F(d)$ as 
a function of the network sizes. 
The insets in~(c) and~(d) are the same as the corresponding panels 
but in double-logarithmic representation [inset (c) only $N=8000$].
For each set $100$~configurations have been averaged.
}
\label{fig:bamexample}
\end{figure}

In order to analyze the correlations in BA model networks 
we generate $100$~configurations, apply the fluctuation analysis, 
and calculate the degree fluctuation function~$F(d)$ among all configurations. 
Figure~\ref{fig:bamexample} shows 
the obtained~$F(d)$ for $m=1,2,3$ and different network sizes
$N=500,1000,\dots,8000$ nodes in semi-logarithmic scale. 
The fluctuation function approximately decays exponentially 
according to $F(d)\sim {\rm e}^{-d/d_\times}$. 
In the case of $m=2$ (Fig.~\ref{fig:bamexample}b) 
straight lines between~$d=2$ and~$d=8$ are found, 
almost parallel for the different sizes.

The dependence on the network size is shown in Fig.~\ref{fig:bamexample}d 
where for $m=2$ the slope~$1/d_\times$ is plotted versus the number of 
nodes~$N$.
In this case, the slope varies between $1/d_\times\approx 0.16$ for 
the small networks and $1/d_\times\approx 0.13$ for the larger ones. 
Further investigations are needed to clarify 
if the slope reaches an asymptotic value.
The BA model has been shown to be uncorrelated 
\cite{PastorSatorrasVV2001,NewmanM2002} and from an analogy with times series analysis, 
one would expect that the fluctuation function decreases according 
to Eq.~(\ref{eq:Fdsimda}) with $\alpha_k=-1/2$ ($\alpha=1/2$).
However, the short distances in the BA model and the exponential decay 
indicate a finite-size effect \cite{LennartzB2009}.

\subsection{Cayley tree}

Since many complex networks are small-world, their diameter 
is rather small and therefore the largest distance~$d$ for 
which the fluctuation function 
can be calculated is also small (see Fig.~\ref{fig:bamexample}a,b).
To overcome this issue we first study a tree structure, 
the Cayley tree at the percolation transition. 
The fractal nature of this system allows us to analyze the fluctuation function 
at longer ranges of $d$.
The Cayley tree~\cite{HavlinN1984,benavrahamH2004} is built by starting from a central node and 
attaching $z$~new nodes to it. In the next generation $z-1$~additional nodes 
are linked to each of the new nodes of the previous generation.
All nodes except the ending ones have degree~$z$.
In percolation, starting from the complete Cayley tree, elements of the tree 
are randomly removed. 
In this case, the percolation transition is at 
$p_{\rm c}=\frac{1}{z-1}$~\cite{benavrahamH2004} and 
the topological dimension of the giant component is 
$d_{\rm f}=2$~\cite{HavlinN1984}.

\begin{figure}
\onefigure[width=0.71\columnwidth]{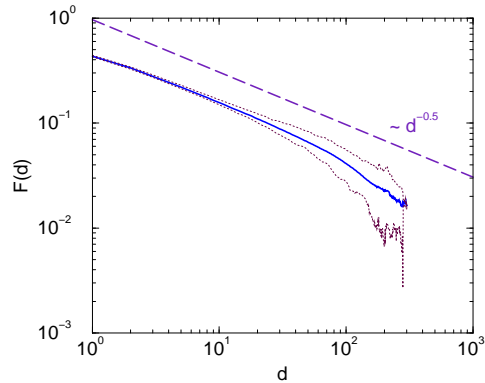}
\caption{
Degree fluctuation function of the $z=3$ Cayley tree at 
percolation transition.
The fluctuation function, $F(d)$, is plotted against the length of the 
shortest paths~$d$. 
The dotted maroon lines represent the quantiles enclosing $90$\% 
of the $100$~configurations (each).
The dashed straight line is a guide to the eye and corresponds to the
exponent $\alpha_k=-1/2$.
}
\label{fig:caylay3}
\end{figure}

Figure~\ref{fig:caylay3} depicts the fluctuation functions~$F(d)$
for the Cayley tree with $z=3$ in generation $n=150$ at the 
percolation transition 
(together with the quantiles enclosing $90$\% of the configurations).
In this case all node degrees are within the range $1\le k\le 3$.
For $2< d < 80$ the fluctuation function follows 
a power-law according to Eq.~(\ref{eq:Fdsimda}) with $\alpha_k\simeq -1/2$ 
($\alpha\simeq 1/2$), indicating uncorrelated degrees of the nodes along 
the paths.
The deviations from the power-law at large distances are due to the finite size 
of the networks.
Since the removal of elements of the Cayley tree happens randomly and 
in particular independent of each other, 
Fig.~\ref{fig:caylay3} is in agreement with what we expect 
($\alpha_k\simeq -1/2$ corresponds to uncorrelated degrees), 
supporting the proposed technique.

\subsection{Fractal network model}

We now use the fractal network model introduced by
Song, Havlin, and Makse \cite{SongHM2006} in which the diameter grows
with the networks size as $N^{1/d_{\rm f}}$.
It has been shown that anti-correlations between node degrees are a 
determinant factor for the fractality of a network 
\cite{SongHM2005,SongHM2006,GallosSHM2007,RozenfeldSM2009,RozenfeldM2009}.

The fractal network is built iteratively 
(an illustration can be found in  Ref.~\cite{RozenfeldM2009}).
It starts in generation~$n=0$ with $2$~nodes connected by $1$~link. 
In generation~$n+1$, $m$~new nodes are attached to the endpoints of 
each link of the previous generation. 
(i) With probability~$1-e$ the corresponding link of generation $n$ is removed and 
$x \le m$ new links are added between the new nodes.
(ii) With probability~$e$ the corresponding link remains and $x-1$ new 
links connect the new nodes.
We use $m=2$, $x=2$, and $n=3,4,5$. In the case of $e=0$, in the first 
generation the network represents a hexagon consisting of $6$~nodes and 
$6$~links. In the second generation, each link of the hexagon is replaced by 
another hexagon, and so on. The larger~$e$, the more short-cuts 
are produced and the more small-world-like the generated network is.
Thus, according to the parameter~$e$, the network is fractal ($e=0$) 
comprising large diameter or non-fractal ($e=1$) being small-world.
On the other hand, although $m$ and $x$ do not alter whether the network is fractal or not,
they are relevant for the number of nodes in the network, for node degrees, and for the number of possible 
paths between pairs of nodes~\cite{SongHM2006}. For example, a large value of $m$ generates a faster increase in
the number of nodes, and a larger $x$ generates more cycles and paths between any pair of nodes.
All three parameters, $m, x, e$ affect the fractal dimension, $d_{\rm f}$, of the network
which is given by~\cite{SongHM2006,RozenfeldM2009}
\begin{equation}
d_{\rm f}=\frac{\ln(2m+x)}{\ln(3-2e)}
\label{eq:fmdf}
\enspace .
\end{equation}
The expression 
diverges when $e = 1$ for a pure small-world network.
For $m=2$, $x=2$ it is $d_{\rm f}=\frac{\ln(6)}{\ln(3-2e)}$. 
The degree distribution is given by 
$p(k)\sim k^{-\left(1+\frac{\ln(2m+x)}{\ln m}\right)}$~\cite{SongHM2006}.

\begin{figure}
\onefigure[width=0.95\columnwidth]{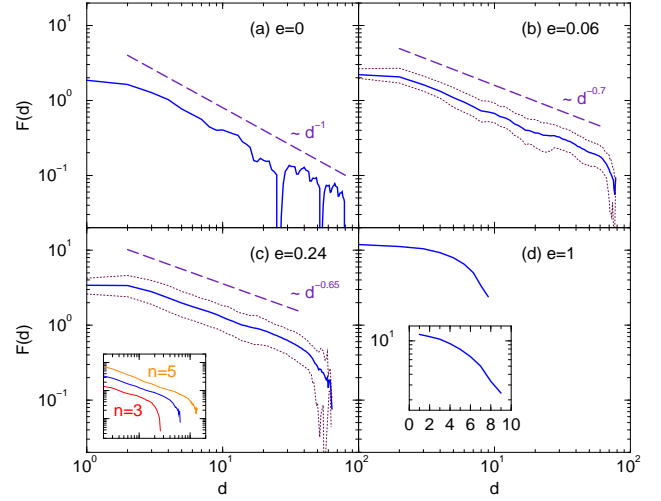}
\caption{
Degree fluctuation functions for the fractal model in 
the fourth generation ($n=4$). 
The fluctuation functions, $F(d)$, are plotted against the length of the 
shortest paths~$d$, for (a)~$e=0$, (b)~$e=0.06$, (c)~$e=0.24$, 
and (d)~$e=1$.
The dotted maroon lines in~(b) and~(d) represent the quantiles 
enclosing $90$\% of the $250$~configurations (each).
The dashed straight lines are guides to the eye and correspond to the
exponents 
(a)~$\alpha_k=-1$,
(b)~$\alpha_k=-0.7$, and
(c)~$\alpha_k=-0.65$.
The cases $e=0$ and $e=1$ are deterministic and accordingly there are 
no uncertainty levels.
Panel~(c) contains an inset with the $F(d)$ for the model with $e=0.24$ 
in third ($n=3$) and fifth ($n=5$, $25$~configurations) generation. 
The inset in panel~(d) is the same as the major panel 
but in linear-log representation.
}
\label{fig:fsexample}
\end{figure}

To analyze this model we generate many configurations 
and apply the fluctuation analysis.
Figure~\ref{fig:fsexample} shows the degree fluctuation functions 
for the model in generation $n=4$ and different values of $e$. 
For $e=0$ (fractal) the fluctuation function decreases as $F(d)\sim d^{-1}$, 
thus $\alpha_k=-1$, whereas $F(d)$ exhibits some structure which is due to 
the deterministic character of the model when $e=0$, 
i.e. for certain distances the sequences of degrees are identical 
leading to vanishing standard deviation among the paths.
In the case of $e=0.06$, the fluctuation function also decreases as 
a power-law but with an exponent $\alpha_k\approx -0.7$. 
Due to finite size effects, $F(d)$ exhibits deviations from a power-law for $d>60$.
For $e=0.24$, Fig.~\ref{fig:fsexample}c, although the network diameter
decreases compared to smaller values of $e$, we still find a power-law 
regime over more than 1 decade. 
The inset of Fig.~\ref{fig:fsexample}c shows the 
fluctuation functions for $n=3$ and $n=5$. 
These power-law decays support the conjecture of long-range correlations.
Since $\alpha=\alpha_k+1<1/2$, the degrees are anti-correlated, 
consistent with previous findings.
The case $e=1$, where the networks is small-world, 
is depicted in Fig.~\ref{fig:fsexample}d. The diameter is much 
smaller and the decay is close to exponential.

\begin{figure}
\onefigure[width=0.7\columnwidth]{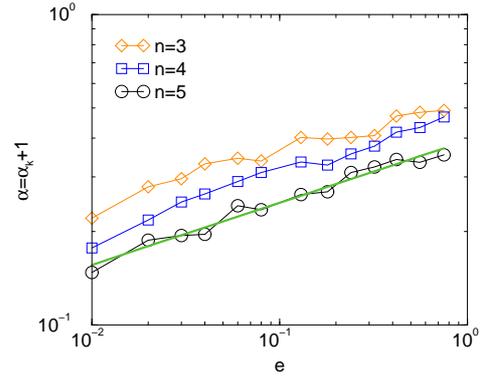}
\caption{
Exponents of the fractal network model. 
The exponents~$\alpha=\alpha_k+1$ are plotted as a function of the 
model parameter~$e$.
The $\alpha_k$-values were obtained from least squares fits to the 
fluctuation functions as exemplified in Fig.~\ref{fig:fsexample}.
We use the model in third, fourth, and fifth generation 
($n=3$, $n=4$: $250$~configurations; $n=5$: $25$~configurations). 
The green straight solid line represents a power-law fit to the exponents 
for~$n=5$ leading to~$\alpha_k+1=\alpha\sim e^{\epsilon}$ 
with~$\epsilon\approx 0.2$.
}
\label{fig:expfit}
\end{figure}

Next, to obtain the exponents~$\alpha_k$ we apply least squares fits to $F(d)$. 
In Fig.~\ref{fig:expfit} the exponents~$\alpha=\alpha_k+1$ are plotted 
against the model parameter~$e$ for the networks in generations~$n=3,4,5$.
We find that the curves of~$\alpha$ vs.~$e$ are parallel for all three values 
of~$n$.
However, the actual value of~$\alpha_k$ systematically decreases with the 
generation~$n$ due to finite-size effects.
Although the range of~$\alpha$ is rather small, we exclude a 
logarithmic dependence of~$\alpha$ on~$e$ since for~$e=0$ we 
find~$\alpha=0$.
Applying the regression $\alpha\sim e^{\epsilon}$ to the $n=5$~exponents 
we find $\epsilon\approx 0.2$.

\subsection{Real-world networks}

The fluctuation analysis serves as a tool to unravel the long-range degree correlations 
present in real-world networks. 
In this section we study the long-range degree correlations in 
$4$~biological networks, 
(i) the Protein Interaction Network of Yeast~\cite{HanBHGBZDWCRV2004}, 
with $777$~nodes and $1797$~links
in which two proteins of the Yeast cell are connected if they interact chemically 
with each other,
(ii) the Metabolic Network of {\it E. coli}~\cite{AlmaasKVOV2004}, 
with $2859$~nodes and $6890$~links, 
in which metabolites within the {\it E. coli} cell
interact biochemically with each other,
(iii) the Human Homology Network~\cite{ArnoldRTTSM2005}, 
with $21710$~nodes and $1289345$~links, 
where proteins of the human cell are connected through a link if
they are biologically homologous (their sequence of amino-acids 
present an E-value smaller than $10^{-30}$)
and
(iv) the Homology Network of 251 Prokaryotic Genomes~\cite{MediniCD2006}, 
with $30727$~nodes and $1206654$~links, 
in which proteins of different organism
are connected through a link if they are biologically homologous.
These 4 networks have been found to exhibit fractality 
\cite{SongHM2005,SongHM2006,GallosSHM2007,GallosSM2008}.

\begin{figure*}
\onefigure[width=0.85\textwidth]{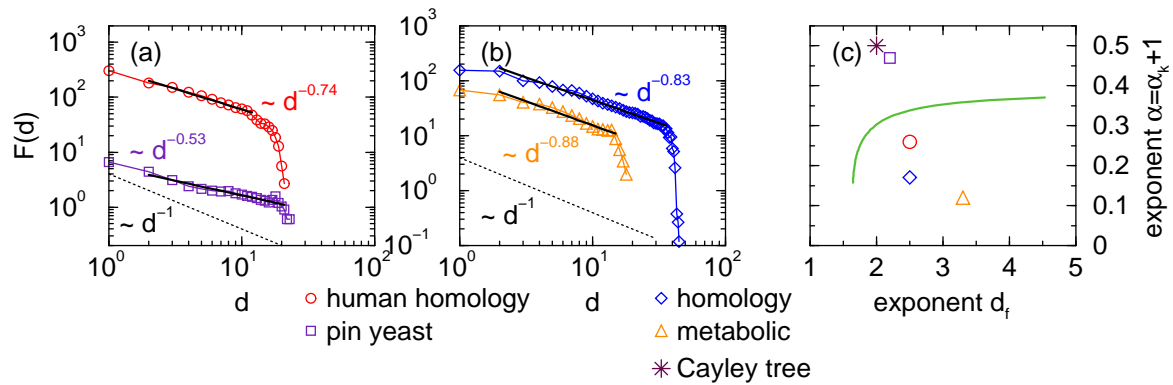}
\caption{
Degree fluctuation functions for real-world networks and comparison of 
the fluctuation exponents with fractal dimension~$d_{\rm f}$.
The fluctuation functions, $F(d)$, are plotted against the length of the 
shortest paths~$d$, for
(a) the human homology network (red circles), 
the pin yeast network (indigo squares), and 
(b) the homology network (blue diamonds), as well as 
the metabolic network (orange triangles).
The black solid straight lines are regressions providing the exponents 
(a) $\alpha_k\simeq -0.74$ (human homology network, $d_{\rm f}\simeq 2.5$), 
$\alpha_k\simeq -0.53$ (pin yeast network, $d_{\rm f}\simeq 2.2$), and 
(b) $\alpha_k\simeq -0.83$ (homology network, $d_{\rm f}\simeq 2.5$), as well as 
$\alpha_k\simeq -0.88$ (metabolic network, $d_{\rm f}\simeq 3.3$).
The black dotted straight lines in the bottom of (a+b) correspond to the 
exponent $\alpha_k=-1$.
(c) Fluctuation exponent $\alpha=\alpha_k+1$ versus 
fractal dimension~$d_{\rm f}$. 
The solid line corresponds to the fit of $\alpha(e)$ from
Fig.~\ref{fig:expfit}, and the asterisk is the corresponding value for 
the Cayley tree.
The values of the real-world networks are given by the corresponding symbols.
}
\label{fig:realnet}
\end{figure*}

Figures~\ref{fig:realnet}a,b show the fluctuation functions obtained for these 
real-world networks. 
The curves exhibit power-law decays over a wide range of~$d$.
For large scales deviations are found that are 
due to the finite size of the networks, similarly to what is seen for the 
fractal network model. 
We obtain the exponents 
$\alpha_k\simeq -0.53$ (pin yeast), 
$\alpha_k\simeq -0.74$ (human homology), 
$\alpha_k\simeq -0.83$ (homology), and 
$\alpha_k\simeq -0.88$ (metabolic).
In Fig.~\ref{fig:realnet}c the corresponding $\alpha=\alpha_k+1$ 
are compared with the fractal dimension values that have been 
found before \cite{SongHM2005,SongHM2006,GallosSHM2007,GallosSM2008}.
The panel also contains the results for the Cayley tree and 
the fit for the fractal model ($n=5$) from Fig.~\ref{fig:expfit} 
when the values $d_{\rm f}$ are calculated according to Eq.~(\ref{eq:fmdf}) 
and with the model parameters ($m=2$, $x=2$, variable~$e$). 

The idea of Fig.~\ref{fig:realnet}c is to compare the fractal dimension with 
the fluctuation exponent. 
Although our results for the real-world networks are scattered in 
Fig.~\ref{fig:realnet}c, they display a decrease in $\alpha$ 
as $d_{\rm f}$ increases. 
The Cayley tree seems to follow the same trend.
In contrast, the simulations in the fractal model reflect the 
opposite. 
This suggests that~$d_{\rm f}$ might not be the only quantity 
related to degree anti-correlations and that the fluctuation exponent provides 
complementary information. 
Further studies, which are beyond the scope of this
work, are necessary to completely determine
the variables affecting the value of $\alpha$.

\section{Discussion and outlook}

In summary, a method for the characterization of spatial degree correlations 
in complex networks is suggested.
The technique is based on fluctuation analysis in analogy to methods used 
in times series analysis.
The fluctuation analysis is applied to the degrees of 
(i) the BA model, 
(ii) the Cayley tree at percolation,
(iii) the fractal network model, and 
(iv) examples of real-world networks.
It is found that fluctuation functions of the BA model decay exponentially.
In contrast, the degree correlations in the fractal network model decay 
as a power-law and according to the obtained exponents the degrees comprise 
long-range anti-correlations. 
Further studies are needed to obtain a better understanding of 
a possible connection between power-law anti-correlations and 
fractality of complex networks.

Such long-range anti-correlations may have structural implications
regarding the stability of networks with respect to random 
removals \cite{CohenEAH2000}.
Long-range correlations could also have an influence on correlations 
between the degree and the betweenness centrality of nodes 
\cite{KitsakHPRPS2007}.
Possible applications include disease spreading, 
see e.g. \cite{KitsakGHLMSM2010} and references therein, 
where it is important whether large degree nodes tend to be linked to other 
large degree nodes or if such hubs are separated by small degree nodes, 
and how correlations at larger distances influence the spreading.
Other applications could be climate networks 
\cite{YamasakiGH2008,TsonisS2008,DongesZMK2009}.

Nevertheless, further research is needed to evaluate the proposed 
fluctuation analysis and the possible dependence of the fluctuation exponents 
on the system size.
In addition, an analytical description of the models is aimed in order 
to gain a better comprehension of long-range correlations in complex networks. 
Furthermore, it would be interesting to vary the model parameters~$m$ and~$x$ 
and to study other fractal networks, as well as more real-world networks. 

Fractality and the small-world property, two seemingly contradicting properties are known 
to coexist in real-world networks~\cite{RozenfeldSM2009}. Typically real-world networks are found
to be fractal up to a certain length scale until the small-world property kicks in at global scale.
Therefore, the method presented in this paper may also be applied to networks that are small-world
and is not limited to fractal networks.

Finally, we would like to point out that our approach can also be extended 
to study other spatial correlations. 
These can be (i) various network properties, as long as they 
can be attributed to the nodes or edges, 
such as clustering, betweenness centrality, and many others, 
or (ii) additional information available in the form of 
characteristic values assigned to the nodes 
(time of addition of the node to the network, activity of the node, etc.), 
or to the links (weight, activity, stability, etc.).
In addition, one could think about including dynamics and analyze 
spatio-temporal correlations.

\section{Acknowledgment}
We wish to thank the Baltic Sea Region Programme (2007-2013) for supporting 
the BaltCICA project.
We thank Jan W. Kantelhardt for useful discussions.

\bibliography{kfa}

\end{document}